\documentclass[aps,pre,preprint,groupedaddress]{revtex4-1}

\usepackage{graphics}
\usepackage{graphicx}
\usepackage{amsfonts}
\usepackage{textcomp}
\usepackage{amssymb}
\usepackage{amsmath}

\usepackage{color}

\begin{document}

\title{Microstructural Degeneracy associated with a Two-Point Correlation Function \\ and its Information Content}

\author{C. J. Gommes$^{1,2}$} \email{cedric.gommes@ulg.ac.be}
\author{Y. Jiao$^2$} \email{yjiao@princeton.edu}
\author{S. Torquato$^{2,3,4}$} \email{torquato@electron.princeton.edu}

\affiliation{$^1$Department of Chemical Engineering, University of Li\`ege, Li\`ege 4000, Belgium}
\affiliation{$^2$Department of Chemistry, Princeton University, Princeton New Jersey 08544, USA}
\affiliation{$^3$Program in Applied and Computational Mathematics,
Princeton University, Princeton New Jersey 08544, USA}
\affiliation{$^4$Princeton Center for Theoretical Science,
Princeton University, Princeton New Jersey 08544, USA}

\date{\today}

\begin{abstract}
A two-point correlation function provides a crucial yet an incomplete characterization of a
 microstructure because distinctly different microstructures may have the same correlation
function. In an earlier Letter [Gommes, Jiao and Torquato, Phys. Rev. Lett. 108, 080601 (2012)],
we addressed the microstructural degeneracy question: What is the number of microstructures
compatible with a specified correlation function? We computed this degeneracy, i.e.,
configurational entropy, in the framework of reconstruction methods, which enabled us
 to map the problem to the determination of ground-state degeneracies. Here, we provide
 a more comprehensive presentation of the methodology and analyses, as well as additional
 results. Since the configuration space of a reconstruction problem is a
hypercube on which a Hamming distance is defined, we can calculate analytically
the energy profile of any reconstruction problem, corresponding to the average energy
of all microstructures at a given Hamming distance from a ground state. The steepness of
the energy profile is a measure of the roughness of the energy landscape associated with
the reconstruction problem, which can be used as a proxy for the ground-state degeneracy.
The relationship between this roughness metric and the ground-state degeneracy is calibrated
using a Monte Carlo algorithm for determining the ground-state degeneracy of a variety of
microstructures, including realizations of hard disks and Poisson point processes at various
densities as well as those with known degeneracies (e.g., single disks of
various sizes and a particular crystalline microstructure). We show that our results can
be expressed in terms of the {\em information content} of the two-point correlation
functions. From this perspective, the {\it a priori} condition for a reconstruction
to be accurate is that the information content, expressed in bits, should be
comparable to the number of pixels in the unknown microstructure. We provide a
formula to calculate the information content of any two-point correlation
function, which makes our results broadly applicable to any field
in which correlation functions are employed.
\end{abstract}


\maketitle

\section{Introduction}

Correlation functions are important structural descriptors that arise in a variety of
disciplines such as solid state physics \cite{Zallen:1983}, signal processing \cite{Proakis:2006},
computer vision\cite{Serra:1982}, statistical physics\cite{Chandler:1987},
geostatistics\cite{Chiles:1999}, and materials science \cite{Matheron:1967,Torquato:2000,Sahimi:2003}.
 Many techniques for structural characterization of materials over a wide range of
 length scales provide data in the form of correlation functions including, but not limited to,
scattering methods \cite{Glatter:1982,Feigin:1987}. Other examples are absorption spectroscopy \cite{Filipponi:1995}, energy transfer analysis \cite{Drake:1991}, nuclear
magnetic resonance \cite{Barral:1992}, and also grey-scale image analysis
\cite{Gommes:2007,Gommes:2010}. Moreover, in the case of {\it in situ} studies
with a nanometer resolution \cite{Svergun:1999,Svergun:2001,Beale:2006, Gommes:2008},
correlation functions are often the only data available experimentally.

Despite the widespread use of correlation functions, the nature of the structural
information they contain remains an open area of active research \cite{Patterson:1944,Yeong:1998,Aubert:2000,Sheehan:2001,Hansen:2005,Fullwood:2008,Jiao:2010A,Jiao:2010B,Berthier:2011}. The central question of the present paper can be put as follows: how accurately is a microstructure
known when the only data available is a two-point correlation function? We shall
focus our analysis on two-phase microstructures and the two-point correlation function $S_2(r)$,
which is defined to be the probability that two random points at a distance $r$ from one
another both belong to the same phase \cite{Torquato:2000}.

Two-point statistics are not sufficient to determine unambiguously a microstructure.
The specification of a given two-point function $S_2(r)$ is equivalent to defining a
macrostate of the system, the degeneracy of which can be expressed as a configurational
entropy. In particular, all space transformations that preserve distances - translation,
rigid rotation and inversion - lead to microstructures with identical two-point statistics.
Following previous work,
we will call such degeneracies  {\em trivial} \cite{Jiao:2010A,Jiao:2010B}.
The focus of the present paper is on {\em non-trivially} degenerate microstructures,
which cannot be obtained from each other through any of the aforementioned
trivial transformations.

Examples of non-trivially degenerate microstructures are Poisson polyhedra tesselations
of three-dimensional space \cite{Serra:1982} and Debye random media
\cite{Debye:1957,Ciccariello:1983,Yeong:1998}, which although having distinct microstructures,
have identical $S_2(r)$. Non-trivial degeneracy is not limited to infinite systems.
 Examples of finite point patterns having the same two-point statistics have been
given as early as 1939, and Patterson coined the word ``homometric'' to qualify them
\cite{Patterson:1939, Patterson:1944}. Very recently, general equations have been
derived that can in principle be solved analytically to obtain homometric microstructures
 \cite{Jiao:2010A,Jiao:2010B}. In the context of crystallography, the degeneracy of the
structural information contained in correlation functions is referred to as the {\it phase problem}.

The phase problem, however, is not universally applicable. A spectacular counterexample
is the so-called direct method of crystallography \cite{Hauptman:1986}, for which
Hauptman and Karle received the 1985 Nobel prize for chemistry. In the field of
computer vision, it has been shown that finite textures are completely characterized
by their orientation-dependent correlation functions \cite{Chubb:2000}.
Many theoretical examples of microstructures with a low degeneracy can be accurately
reconstructed from their correlation function alone
\cite{Rozman:2001,Rozman:2002,Hansen:2005,Fullwood:2008}. All these
examples have in common that they incorporate orientation information
\footnote{In the case of the direct method of crystallography, the absolute
orientation is generally lost through the measurement of a powder scattering pattern.
The orientation with respect to the unit cell, however,
is known from the indexing of the Bragg peaks.}.
The focus of the present work is on radial correlation functions in
which orientation information is averaged out. This simplification is
 relevant to many experimental contexts, notably small-angle scattering
\cite{Glatter:1982,Feigin:1987}, where the correlation function is
generally rotationally averaged through the measurement of powder
scattering patterns, as well as to isotropic disordered systems in general \cite{Torquato:2000}.

The understanding of the structural information in radial correlation
functions has been considerably advanced through the use of reconstruction
algorithms, which aim at producing microstructures with a specified
correlation function via the minimization of a prescribed energy
functional \cite{Gagalowicz:1981,Rintoul:1997,Yeong:1998,Aubert:2000,Jiao:2009}.
In the case of a reconstruction based on two-point correlation functions,
a natural choice for the energy functional is \cite{Rintoul:1997,Yeong:1998}
\begin{equation} \label{eq:definition_E}
E = \sum_r \left[ \hat S_2(r) - S_2(r) \right]^2 \ ,
\end{equation}
where $\hat S_2(r)$ is the target two-point correlation function, $S_2(r)$
is the correlation function of the microstructure, i.e., the configuration
being optimized, and the sum is over all measurable distances.
This definition of the energy is equivalent to a norm-2 error:
it is non-negative and it vanishes only for those configurations that
satisfy $S_2(r) = \hat S_2(r)$. In this context, the question of
the degeneracy associated with a given correlation function is
equivalent to determining the number of microstructures having zero
energy, i.e., the {\it ground-state degeneracy} of the energy functional \cite{Gommes:2012}.

The minimization of Eq. (\ref{eq:definition_E}) is generally done by
discretizing the microstructure on a
grid with periodic boundary conditions, and by using either a steepest
 descent \cite{Gagalowicz:1981,Rozman:2001} or a simulated annealing
\cite{Yeong:1998} algorithm. In the case of a two-phase microstructure,
 which can be thought of as an image with black and white pixels, the
simulated annealing reconstruction proceeds as follows. Starting from
any configuration, with value $E_i$ of the energy functional
Eq. (\ref{eq:definition_E}), a black pixel is chosen randomly and moved
to any available white position. The function $S_2(r)$ is updated and
the new energy $E_j$ is calculated. The move is accepted with probability
\begin{equation} \label{eq:Metropolis}
p_{i \to j} = \min \left\{1, \frac{\exp(E_i/T)}{\exp(E_j/T)} \right\} \ ,
\end{equation}
where $T$ is a temperature parameter \cite{Kirkpatrick:1983}.
All energy-decreasing moves are therefore accepted but some
energy-increasing moves are accepted as well, depending on the
chosen temperature. The latter moves are necessary to ensure that
the entire configuration space be explored in principle, and that
the system is not trapped in a local minimum of $E$. Simulated
annealing algorithms consist in starting at a high temperature,
and progressively decreasing the temperature until convergence
is reached ($E \approx 0$). This type of approach has been widely
used for microstructure reconstruction, in the context of
both applications \cite{Alexander:2009, Kim:2009, Schlueter:2011,Grew:2012}
and theoretical investigations \cite{Yeong:1998,Sheehan:2001,Rozman:2002,Jiao:2007}.
The latter include generalizations to
other types of statistical microstructure descriptors besides
$S_2(r)$ \cite{Rintoul:1997}, most notably to higher-order
correlation functions \cite{Aubert:2000, Jiao:2009} as well as to
cluster correlation functions \cite{Jiao:2009,Zachary:2011}.

\begin{figure}
\begin{center}
\includegraphics[width=7cm, keepaspectratio]{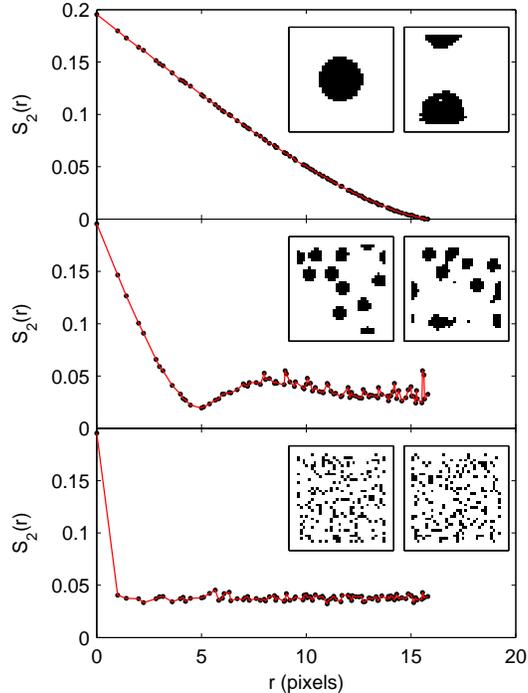}
\end{center}
\caption{(Color online) From top to bottom: reconstructions of a
single disk, hard disks, and the realization of a Poisson point
process under periodic boundary conditions. In each case,
the target (left) and reconstructed (right) microstructures
are shown. The target ($\bullet$) and reconstructed ($-$)
correlation functions are indistinguishable on the scale of the figure.
The size of the grid is $32 \times 32$ pixels with $N_1 =200$.
These examples strongly suggest that the two-point function of a
single sphere under periodic boundary conditions is only trivially
degenerate through translation, but that the two-point
degeneracy of a Poisson point process has a large non-trivial contribution.}
\label{fig:REC_examples}
\end{figure}

Examples of reconstructions of two-phase microstructures under
periodic boundary conditions are given in Fig. \ref{fig:REC_examples}.
In the case of the single disk, the reconstructed microstructure
is almost identical to the target, except for a translation
(top portion of Fig. \ref{fig:REC_examples}). In the case of the
reconstruction of the hard disks (middle portion of
Fig. \ref{fig:REC_examples}), the characteristic size of the disks
as well as the average distance between them is recovered.
However, an exact reconstruction of the target configuration is
not possible; spurious objects are also formed through the partial
 merging of neighboring disks. In the case of a realization of a
Poisson point process (randomly coloring a pixel black according
to a prescribed volume fraction), the reconstructed and the target
microstructures might look superficially similar because they
both appear to be random distributions of black pixels (bottom
portion of Fig. \ref{fig:REC_examples}). However, the two
microstructures have very little in common if one is interested
in the exact configurations of the pixels, although an excellent
match is obtained between $S_2(r)$ and $\hat S_2(r)$.
This illustrates the concept of non-trivial degeneracy \cite{Patterson:1939,Jiao:2010A,Jiao:2010B}.

In a recent Letter \cite{Gommes:2012}, we presented a general
theoretical framework for estimating quantitatively the structural
degeneracy corresponding to any specified correlation function.
This was achieved by mapping the problem to the estimation of a
ground-state degeneracy through the use of Eq. (\ref{eq:definition_E}).
Here we provide a more comprehensive presentation of
the methodology and analyses, including a quantitative characterization
of the energy landscape associated with the reconstruction as
well as a detailed derivation of the degeneracy metric.
Moreover, we show that our results can be expressed in terms of
the {\em information content} of the two-point correlation functions.
Although the present work focuses on two-dimensional media in
Euclidean space, our procedure can be applied in any space
dimension and generalized to non-Euclidean spaces (e.g., compact and hyperbolic spaces).

The remainder of the paper is organized as follows.
In Sec. II, we discuss the degeneracy of the two-point statistics
for a variety of microstructures that are used as benchmarks
throughout the rest of the paper. We consider successively small
systems - for which all the configurations can be enumerated -
intermediate systems - for which the degeneracy can be
determined via a Monte Carlo method we presented elsewhere
\cite{Gommes:2012} - and large systems for which neither of
the aforementioned two methods apply and one needs to use the
reconstruction method. In Sec. III, we devise an analytical method,
based on a random walk in configuration space, to characterize
the energy landscape associated with reconstruction. In particular,
we determine a characteristic energy profile for the basin
of each ground state. In Sec. IV, we show that the ground-state
degeneracy of reconstruction problems is related to the roughness
of the energy landscape. We introduce a roughness metric that
can be calculated from $\hat S_2(r)$ alone, and we show
definitively that it is correlated with the microstructure degeneracy.
In Sec. V, the degeneracy is expressed in terms of the {\it information content}
of $\hat S_2(r)$, and a formula is proposed relating the
roughness metric to this information content. The practical usefulness of our results is discussed.

\section{The degeneracy of two-point statistics}

\subsection{Small-system-size microstructures: countable examples}
\label{sec:small}

\begin{figure}
\begin{center}
\includegraphics[width=7.5cm, keepaspectratio]{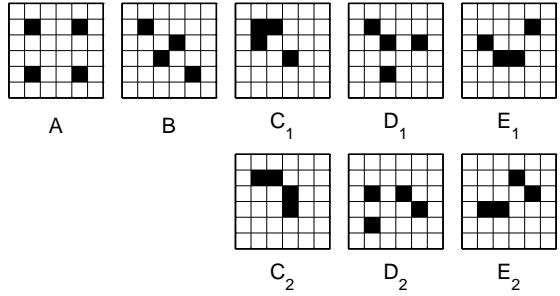}
\end{center}
\caption{Examples of small-system-size microstructures
(under periodic boundary conditions) having different two-point
degeneracies. From A to E, the degeneracies are $\Omega_0 = $
$N$, $2 N$, $8N$, $12 N$, $16 N$, with $N$ the total number of
pixels in the grid (see text). Systems C to E have a non-trivial contribution to their degeneracy.}
\label{fig:Degenerate}
\end{figure}

The present paper is restricted to two-phase digitized microstructures,
which can be thought of as images with $N_1$ black pixels and
$N_0=N-N_1$ white pixels. However, our analysis can be easily
generalized to multiphase microstructures. We shall first
consider the very small microstructures of
Fig. \ref{fig:Degenerate} with $N_1=4$. They will be
analyzed in some detail and will serve as a benchmark for
analytical methods applicable to larger and more complex microstructures.

For any finite microstructure it is always possible to refer to
the pixels through a linear index, $i = 1$ to $N$, independently
of the actual dimensionality. A finite microstructure is therefore
completely characterized by a $N$-dimensional vector,
with components $I(i)$ equal to $1$ when point $i$ is a black pixel,
and $0$ otherwise. The two-point correlation function $S_2(r)$ of
the black phase is defined as the probability that two random
pixels at a distance $r$ from one another are both black \cite{Torquato:2000}.
This can be written formally as
\begin{equation} \label{eq:definition_P}
S_2(r) = \frac{1}{N \omega_r} \sum_{i=1}^N \sum_{j=1}^N I(i) I(j) D_r(i,j) \ ,
\end{equation}
where $D_r(i,j)$ takes the value $1$ if the distance between pixels
$i$ and $j$ is $r$, and $0$ otherwise. The quantity $\omega_r$ is
defined as $\omega_r=\sum_i D_r(i,j)$.
In Eq. (\ref{eq:definition_P})
the double sum counts the pairs of black pixels separated by a distance $r$,
and the pre-factor normalizes that count by the total number of pixel
pairs at a distance $r$ from one another. The periodic boundary
conditions are incorporated in the definition of the operator $D_r(i,j)$.
We assume in the rest of the paper that the discretizing grid is uniform in the sense that $\omega_r$ is independent of $j$.

The use of a discrete pixel grid is equivalent to a ``quantizer" problem \cite{Torquato:2010}, in which every point of the microstructure is quantized to the centroid of its closest pixel. The distances $r$ between pairs of points are therefore approximated by distances that are compatible with the grid. A square grid is used throughout the present paper. For finite-size systems, the quantization naturally introduces some grid-specific artifacts \cite{Jiao:2008}. However, the quantization error decreases and becomes zero in the limit of infinitely large microstructures.

The two-point correlation functions of the microstructures of
Fig. \ref{fig:Degenerate} are given in Table \ref{tab1} under the form
$ P(r) = N \omega_r S_2(r)/2$. The quantity $P(r)$ is equal to the
number of pairs of points at distance $r$ from one another.
Note that although configurations $C_1$ and $C_2$ are different,
they have identical two-point characteristics.
The same applies to $D_1$ and $D_2$, as well as to $E_1$ and $E_2$.
A complete enumeration of all microstructures with $N_1 = 4$
shows that there is no other configuration with the same $S_2(r)$.

\begin{table*}
\caption{\label{tab1} Number of pairs $P(r) = N \omega_r S_2(r)/2$
in microstructures A to E of Fig. \ref{fig:Degenerate}. Note
that $P(r)$ is identical for configurations $C_1$ and $C_2$, $D_1$ and $D_2$, $E_1$ and $E_2$.}
\begin{ruledtabular}
\begin{tabular}{lccccccccc}
$r$ & 1 & $\sqrt{2}$ & 2 & $\sqrt{5}$ & $\sqrt{8}$ & 3 & $\sqrt{10}$ & $\sqrt{13}$ & $\sqrt{18}$ \\
\hline
$ P_A(r)$  &  0  &  0  &  0  &  0  &  0  &  4  &  0  &  0  &  2  \cr
$ P_B(r)$  &  0  &  1  &  0  &  4  &  0  &  0  &  0  &  0  &  1  \cr
$ P_C(r)$  &  2  &  1  &  0  &  2  &  1  &  0  &  0  &  0  &  0  \cr
$ P_D(r)$  &  0  &  1  &  2  &  0  &  1  &  0  &  2  &  0  &  0  \cr
$ P_E(r)$  &  1  &  1  &  0  &  2  &  1  &  0  &  1  &  0  &  0  \cr
\end{tabular}
\end{ruledtabular}
\end{table*}

Configuration $A$ in Fig. \ref{fig:Degenerate} is uniquely defined
by its two-point function, and therefore is only trivially degenerate.
On grids with $N$ points the total number of translations is $N$;
the number of rotations is $1$, $2$ or $4$, depending on the rotational
symmetry of the configuration; and the number of mirror configurations
is $1$ or $2$, depending on its chirality. Due to the symmetry and
chirality of configuration A, only translation contributes to its
degeneracy, which is therefore $\Omega_0 = N$. In the case of
configuration $B$, the two possible orientations contribute an extra factor 2, i.e. $\Omega_0 = 2 N$.

Configurations $C_1$ and $C_2$ are the "Kite \& Trapezoid"
examples discussed in Refs. \cite{Patterson:1944,Jiao:2010A,Jiao:2010B},
which are non-trivially degenerate. In this case, $\Omega_0 =2 \times 4 \times N $,
where the factor 2 is the non-trivial contribution, and the factor
4 accounts for the possible orientations.

Configurations $D_1$ and $D_2$ are also non-trivially degenerate.
Configuration $D_2$ is, however, chiral so it has to be counted twice.
This leads to $\Omega_0 = (1 + 2) \times 4 \times N$. Finally,
non-trivially degenerate configurations $E_1$ and $E_2$ are both chiral.
This leads to $\Omega_0 = (2 + 2) \times 4 \times N$.

\subsection{Intermediate-system-size microstructures: Monte Carlo analysis}
\label{sec:MC}

The complete enumeration of degenerate microstructures is intractable
for systems even barely larger than those represented in
Fig. \ref{fig:Degenerate}. In the present section, we discuss a
Monte Carlo (MC) algorithm for estimating $\Omega_0$,
which we introduced previously \cite{Gommes:2012}. It can be applied to larger systems.

The approach is based on a general MC algorithm for estimating
the density of states (DOS) developed by Wang and Landau
\cite{Wang:2001A,Wang:2001B} and further improved and analyzed by others
 \cite{Dayal:2004,Zhou:2005,Belardinelli:2007}. The algorithm has been
 applied to a host of problems in condensed matter physics \cite{Yamaguchi:2001},
in biophysics \cite{Rathore:2002}, and in logic \cite{Ermon:2010}.
The DOS $\Omega(E)$ is defined as the number of states having
energy equal to $E$. The logarithm of $\Omega(E)$ is equal to the
entropy calculated in the microcanonical ensemble
associated with Eq. (\ref{eq:definition_E}).
The ground-state degeneracy $\Omega_0$ is the value taken by $\Omega(E)$ for $E = 0$.

A canonical Monte Carlo simulation with transition probability
given by Eq. (\ref{eq:Metropolis}) leads the system to visit
any energy with a probability $p(E) \sim \exp(-E/T)$ \cite{Metropolis:1953}.
The algorithm of Wang and Landau is based on the
observation that a transition probability of the form
\begin{equation} \label{eq:Wang_Landau}
p_{i \to j} = \min \left\{1, \frac{\Omega(E_i)}{\Omega(E_j)} \right\}
\end{equation}
would lead the system to visit all energies with equal probability.
The density of states $\Omega(E)$ is, however, unknown so the algorithm is iterative.

The starting value is set to $\Omega(E) = 1$ for all energies,
and the system is let evolve according to Eq. (\ref{eq:Wang_Landau}),
while updating an histogram $H(E)$. Each time an energy is visited
the corresponding bin is updated, $H(E) \to H(E) +1$, and the
estimated density of states is updated according to
$\Omega(E) \to F \times \Omega(E)$ where $F$ is a numerical
factor larger than 1. The evolution continues according to
Eq. (\ref{eq:Wang_Landau}) with the updated value of $\Omega(E)$.
The evolution is stopped when a flat histogram is obtained.
At this point, the histogram $H(E)$ is reset to $0$, $F$ is
reduced to a value closer to 1, and the evolution starts over again.
The entire procedure is repeated until $F$ becomes lower than
a prescribed accuracy. Algorithmic details are provided in
 the Supplementary Material \cite{SupportingInformation}.

\begin{figure}
\begin{center}
\includegraphics[width=7cm, keepaspectratio]{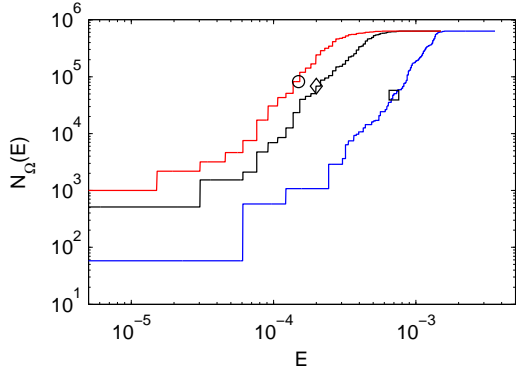}
\end{center}
\caption{(Color online) Cumulative DOS associated with the
 reconstruction of configurations A ($\square$), C ($\diamond$), and E ($\circ$) of Fig. \ref{fig:Degenerate}.}
\label{fig:DOS_4points}
\end{figure}

The accuracy of the MC algorithm was tested by applying it to
the microstructures of Fig. \ref{fig:Degenerate}. The results
 are plotted in Fig. \ref{fig:DOS_4points} in the form of cumulative DOS
\begin{equation}
N_\Omega(E) = \sum_{e\leq E} \Omega(e) \ .
\end{equation}
The MC algorithm provides $\Omega(E)$ only to within an unknown
multiplicative constant, which is determined by imposing
$ \sum_E \Omega(E) $ to be equal to the total number of
 configurations $\Omega_{tot}$. The latter is equal to the
 number of different ways in which $N_1$ black pixels can be
 chosen among a total of $N$ possible pixels, i.e.
\begin{equation}
\Omega_{tot} = \binom{N}{N_1} \ .
\end{equation}
The cumulative DOS plotted in Fig. \ref{fig:DOS_4points}
satisfies $N_\Omega(E \to \infty) = \Omega_{tot}$ and $N_\Omega(E \to 0) = \Omega_0$.

Three independent MC estimations have been calculated for each
 microstructure in Fig. \ref{fig:Degenerate}, yielding three
independent estimates of $\Omega_0$. The results are: $66 \pm 7 $
 for configuration A compared to the exact value $64$; $140 \pm 6$
 for configuration B compared to $128$; $500 \pm 68$ for
configuration C compared to $512$; $769 \pm 18 $ for configuration
D compared to $768$; and $ 991 \pm 85 $ for configuration
$E$ compared to $1024$. The exact values are those calculated
in Sec. \ref{sec:small} with $N=64$. The agreement with the MC estimates is excellent.

\begin{figure}
\begin{center}
\includegraphics[width=7cm, keepaspectratio]{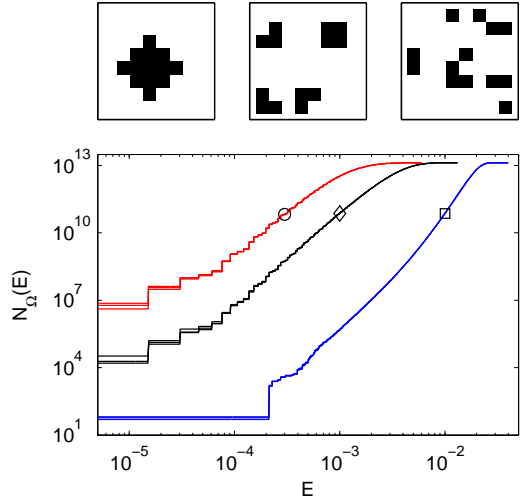}
\end{center}
\caption{(Color online) Top, from left to right: single disk,
hard disks, and Poisson point process realization; Bottom:
cumulative DOS associated with their reconstruction from
 $S_2(r)$ in the case of the disk ($\square$), hard disks
($\diamond$), and the Poisson point process($\circ$).}
\label{fig:DOS_13points}
\end{figure}

Figure \ref{fig:DOS_13points} shows MC estimates of the density
 of states for larger microstructures, with $N_1 = 13$ on a
 $8 \times 8$ grid. The microstructures are qualitatively
similar to those of Fig. \ref{fig:REC_examples}, namely a
single disk, hard disks, and a Poisson point process, all
under periodic boundary conditions. In the case of a single
 disk, the MC estimation provides the value $\Omega_0 = 58 \pm 8$,
corresponding to the $64$ possible translations.
This confirms that the disk is only trivially degenerate.
By contrast, the value found for the Poisson point
process is $\Omega_0 = (11 \pm 1) \ 10^6 $, which is
orders of magnitude larger than any possible trivial
 contribution from translation and rotation. In the
case of the hard disks, we find $\Omega_0 = (23 \pm 4) \ 10^3$.

\subsection{Large-system-size microstructures}
\label{sec:large}

The MC algorithm does not converge for systems larger than
about $10 \times 10$ pixels. With larger systems the criterion
for flat histograms is rarely reached, even with as many
 as $10^9$ MC steps. Moreover, when flat histograms are
 indeed obtained, the estimated value of $\Omega_0$ is much
smaller than 1, which shows that the algorithm explores
only a small fraction of the complete configuration space.
These numerical difficulties are consistent with previous
 observations that flat-histogram algorithms have a
 convergence time that increases exponentially with system size \cite{Dayal:2004}.

It is therefore difficult to estimate the $S_2$-degeneracy
of systems as large as the one shown in Fig. \ref{fig:REC_examples},
except in the particular case where the microstructure
is only trivially degenerate.
It has to be stressed that reconstructing exactly a
 degenerate microstructure is very unlikely. Therefore,
whenever a reconstruction leads to a translated, rotated,
and inverterted version of the target, this can be
 considered as very strong evidence that the microstructure
is only trivially degenerate. In the remainder of
the paper, we shall refer to a microstructure as
being {\em non-degenerate}, whenever it has only a trivial degeneracy.

In continuous Euclidean space under periodic boundary
 conditions, an example of non-degenerate microstructure
 is provided by the single sphere (composed of a
large number of pixels). This results from the observation
 that $S_2(0)$ is equal to volume fraction of the solid
phase and that the negative slope of $S_2(r)$ for $r \to 0$
is proportional to its surface area \cite{Debye:1957}.
 A sphere is non-degenerate because it is the microstructure
that realizes the lowest possible surface area for a
 given volume fraction: the two-point correlation
function of any microstructure other than a single
sphere would have a larger slope at the origin, which
 would result in a positive energy according to
Eq. (\ref{eq:definition_E}).

This observation can be expressed in a way that
generalizes to discrete microstructures: for a given number
 of black pixels $N_1$, a single sphere is non-degenerate
 because it is the microstructure that realizes the
largest value of $S_2(\epsilon)$, where $\epsilon$
is a very small distance. Similarly, any configuration
 with $N_1 = 13$ other than the disk of
 Fig. \ref{fig:DOS_13points} has a smaller value of
$P(\sqrt{2})$, where it is to be recalled that $P(r)$
is the number of pairs of points with distance $r$.
The same applies to configuration $A$ of
Fig. \ref{fig:Degenerate}, which is not a disk:
that particular microstructure is non-degenerate
because any other configuration with $N_1 = 4$ has
a smaller value of $P(3)$. The origin of the degeneracy
of hard-disk systems is touched on in Sec. VI.

\begin{figure}
\begin{center}
\includegraphics[width=7cm, keepaspectratio]{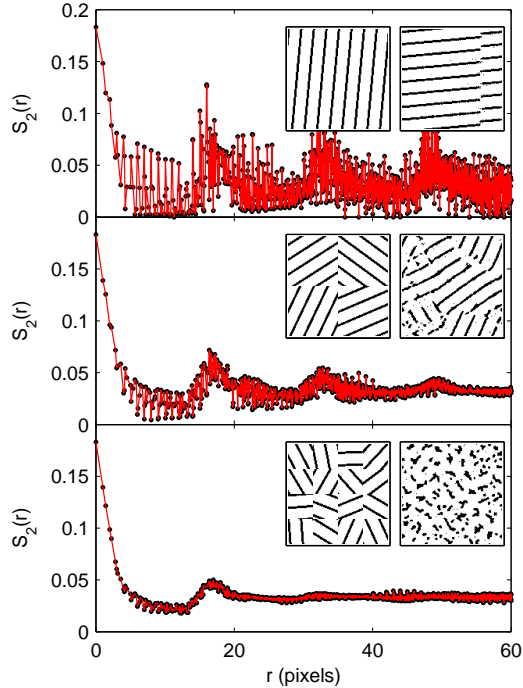}
\end{center}
\caption{(Color online)  From top to bottom: reconstructions
 of a crystal and of two polycrystals with decreasing
crystallite sizes. In each case, the target (left) and
reconstructed (right) microstructures are shown.
 The target ($\bullet$) and reconstructed ($-$)
correlation functions are indistinguishable on the
 scale of the figure. The size of the grid is
$128 \times 128$ pixels under periodic boundary
conditions, and $N_1 =3000$.}
\label{fig:REC_crystal}
\end{figure}

The analysis of non-degeneracy in terms of extremal
values of $P(r)$ leads to non-intuitive results. When
microstructures are defined on a grid, distances and
orientations are not independent: for instance, a pair
of points at a distance $\sqrt{8}$ from one another is
necessarily oriented at 45\textdegree \ with respect
to both axes. A very anisotropic microstructure such
as the crystal on the top of Fig. \ref{fig:REC_crystal}
minimizes $P(r)$ for a set of distances corresponding to
orientations orthogonal to the stripes. The figure clearly
shows that $S_2(r)$ vanishes for a set of well-defined distances.
It should therefore not be surprising that such a highly
anisotropic microstructure is non-degenerate.
The non-degeneracy of the crystal is confirmed by the
fact that the reconstructed microstructure in
Fig. \ref{fig:REC_crystal} is a translated and rotated
copy of the target. The vertical discontinuity in the
middle of the reconstruction results simply from the
target not having the same periodicity as the box.

When a large crystal in a periodic box is split into a
collection of randomly oriented smaller crystallites
(Fig. \ref{fig:REC_crystal} middle and bottom rows),
its anisotropy is reduced and there are no longer values
of $r$ at which $P(r)$ is extremal. Accordingly
the reconstruction becomes less accurate, which means
that the microstructure becomes more degenerate. A more
quantitative analysis of this issue is provided in Sec. V.

\section{Characterization of the Energy Landscape}

\subsection{The Structure of Configuration Space $\mathcal{C}$}

\begin{figure}
\begin{center}
\includegraphics[width=4.5cm, keepaspectratio]{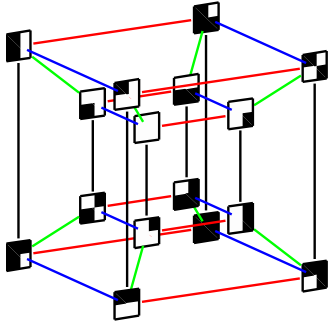} \\
\end{center}
\caption{(Color online) The configuration space $\mathcal{C}$
of a two-phase microstructure is an $N$-dimensional hypercube
on which Hamming distance can be defined. Any move along a
$N$-dimensional direction corresponds to changing the color
of a particular pixel. In the case of a $2 \times 2$
microstructure the configuration space is a tesseract,
with the fourth dimension represented as the edges
joining the outer and inner cubes (corresponding to the upper-left pixel).}
\label{fig:Hypercube}
\end{figure}

The complete configuration space $\mathcal{C}$ of two-phase
 microstructures with $N$ pixels is the set of vertices of
an $N$-dimensional hypercube \cite{Gommes:2012}. This
results from the properties of the indicator vector,
$I(i)$, which can take only values $0$ and $1$. Moving
along a given $N$-dimensional direction (along an edge of
 the hypercube) is equivalent to interchanging a white (black)
with a black (white) pixel. In the example of Fig. \ref{fig:Hypercube},
any movement along the fourth dimension (joining the outer and inner cubes)
corresponds to changing the upper-left pixel.

In the situation relevant to reconstruction, not all the
vertices of the hypercube are accessible because the number of black pixels is kept constant, i.e.
\begin{equation} \label{eq:hyperplane}
\sum_{i=1}^N I(i) = N_1 \ ,
\end{equation}
which means that all realizable microstructures lie on
the intersection of the hypercube with a hyperplane.
Once a target correlation function $\hat S_2(r)$ is specified,
each vertex is assigned an energy through Eq. (\ref{eq:definition_E}).

What we refer to as the energy landscape is the set of
values taken by the energy functional $E$ on the vertices
of the $N$-dimensional hyperplane. A reconstruction consists
in exploring the energy landscape until a vertex is
found with $E=0$. The DOS $\Omega(E)$ determined in section
\ref{sec:MC} is the number of vertices having a given
energy $E$. The problem we address in this section is
that of the spatial variability of $E$ in configuration
space $\mathcal{C}$. This analysis is motivated by the
observation, in many fields of physics, that systems with
large ground-state degeneracies generally have a
rough energy landscape \cite{Wales:1998,Debenedetti:2001}.
If we can characterize the roughness of the energy
landscape in terms of $\hat S_2(r)$ we can estimate
the ground-state degeneracy $\Omega_0$.

In order to characterize the spatial variability of
$E$ in configuration space, it is necessary to define
a distance. A natural choice is the {\it Hamming} distance,
which counts the number of edges between any two vertices.
The Hamming distance within the hyperplane defined by
Eq. (\ref{eq:hyperplane}) takes only even values.
 The distance $d[A,B]$ between two microstructures
$A$ and $B$ is therefore defined as half the Hamming
 distance
\begin{equation}  \label{eq:definition_d}
d[A,B]=\frac{1}{2} \sum_{i=1}^N \left| I_A(i)-I_B(i) \right| \ ,
\end{equation}
where $I_A(i)$ and $I_B(i)$ are the indicator vectors.
In real space, this distance $d$ is the smallest number
of Monte-Carlo-like black pixel displacements that
are required to pass from $A$ to $B$. The largest
possible distance is $d = N_1$ when the two
microstructures have no pixel in common
\footnote{We assume throughout the paper that
black pixels are fewer than white pixels, which
does not limit the generality of the analysis.}.

\subsection{Exploring the Energy Landscape with a Random Walk}

\begin{figure*}
\begin{center}
\includegraphics[width=12cm, keepaspectratio]{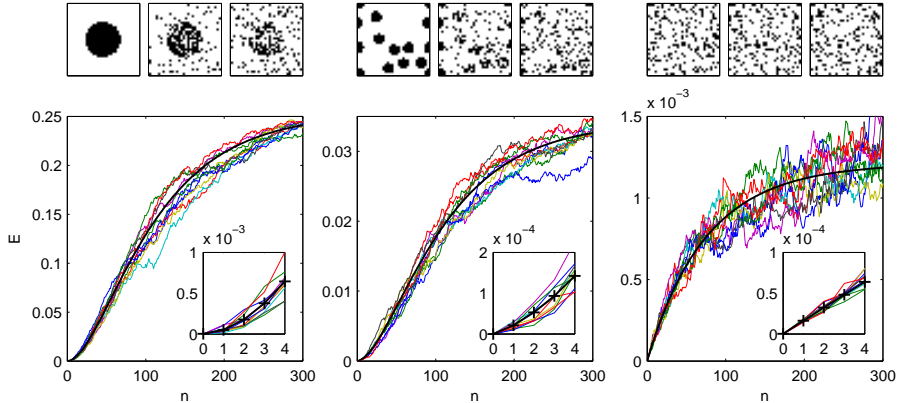} \\
\end{center}
\caption{(Color online) Random walks in the configuration space
$\mathcal C$ of a single (discretized) disk, hard disks, and a
Poisson point process (left to right), all with $N_1 = 200$ pixels
under periodic boundary conditions. The irregular curves are the
energies visited during particular realizations of the random walk,
and the thick black line is the average value calculated
analytically through Eq. (\ref{eq:E_n_average}). The microstructures
shown on top are the starting ground states and the configurations
reached after $n=80$ and $n=160$ random moves.}
\label{fig:Random_Walk}
\end{figure*}

The energy landscape can be characterized
analytically through a random walk in configuration space.
This is illustrated in Fig. \ref{fig:Random_Walk}.
Starting from a ground state of a reconstruction problem,
with $\hat S_2(r) = S_2(r)$, the system moves randomly to
any configuration at Hamming distance $d=1$ from the
current configuration. This is equivalent to a standard
Metropolis Monte Carlo with $T \to \infty$ [see Eq. (\ref{eq:Metropolis})].
When the number of moves  $n$ increases, the random walk explores
the configuration space $\mathcal{C}$ over increasingly large
Hamming distances $d$ from the starting ground state.

The rate at which the average energy $\left< E \right>^{(n)}$ visited
by the random walk increases with the number $n$ of moves characterizes
the energy landscape of a given reconstruction problem.
In the examples of Fig. \ref{fig:Random_Walk}, the energy curve
of the Poisson point process is steeper than that of the single disk,
which suggests smaller basins. We now proceed to analytically
calculate the values of $\left< E \right>^{(n)}$ as a
function of the characteristics of the starting ground state.

The only {\it a priori} information about the ground states of a
given reconstruction problem is their one-point and two-point
characteristics: $\phi=N_1/N$ and $\hat S_2(r)$. The other
characteristics, in particular the higher-order correlation functions,
may differ significantly from one ground state to another.
Let us assume for now that the starting ground state of the random
walk is perfectly known through its indicator vector $I(i)$.

Instead of using $S_2(r)$, it is convenient to use the equivalent
autocovariance $\chi(r)$ defined as
\begin{equation}
\chi(r) = S_2(r) - \phi^2 \ ,
\end{equation}
where $\phi = N_1/N$ is the fraction of black pixels.
The average energy after $n$ random moves can be written in terms of $\chi(r)$ as
\begin{equation} \label{eq:E_n_average_general}
\left< E  \right>^{(n)} = \sum_r \hat \chi^2(r) + \big< \chi^2(r)  \big>^{(n)}
- 2 \hat \chi (r) \big< \chi(r)  \big>^{(n)} \ ,
\end{equation}
where we have used the notations $\hat \chi (r) = \hat S_2(r) -\phi^2$,
which is associated with the ground state, and $\left< . \right>^{(n)}$
for any average value at step $n$. At each step of the random walk
there are $N_0 N_1$ possible moves, so that the total number of
possible walks of length $n$ is $(N_0 N_1)^n$: the averages $<.>^{(n)}$
are calculated over all these possible paths. We now calculate
successively $\left< \chi(r) \right>^{(n)}$ and $\left< \chi^2(r) \right>^{(n)}$,
which are required to calculate and $\left< E \right>^{(n)}$ through Eq. (\ref{eq:E_n_average_general}).

When a black pixel $p$ is moved to position $q$ previously
occupied by a white pixel, the indicator vector becomes
\begin{equation} \label{eq:I_prime}
I'(i)=I(i) + \delta(i,q) -  \delta(i,p) \ ,
\end{equation}
where $\delta(i,q)$ is the Kronecker delta function. Using
the definition of $S_2(r)$, Eq. (\ref{eq:definition_P}),
the autocovariance is then found to become
\begin{eqnarray} \label{eq:P_prime}
\chi'(r;p,q)  =  \chi(r)  + \frac{2}{N \omega_r}  \Big\{ \delta(r,0) - D_r(p,q) \nonumber \\
+ \sum_i I(i) [D_r(q,i)-D_r(p,i)]  \Big\}
\end{eqnarray}
for that particular move. The average value of $\chi'(r;p,q)$
is then simply calculated as
\begin{equation} \label{eq:P_average}
\left< \chi'(r) \right >  =   \frac{1}{N_0 N_1} \sum_{p \ q} I(p) \left( 1-I(q) \right) \chi'(r;p,q) \ ,
\end{equation}
where the factor $I(p) \left( 1-I(q) \right)/(N_0 N_1)$ accounts
for the fact that $p$ and $q$ are uniformly distributed over the
black and white phases, respectively.

Combining Eqs. (\ref{eq:P_prime}) and (\ref{eq:P_average}),
the average autocovariance $\chi(r)$ after a single random move is found to be
\begin{equation} \label{eq:Evolution_chi}
\left< \chi'(r) \right >  = \alpha \ \chi(r) + O(N^{-2})
\end{equation}
with $ \alpha = 1 - 2 N / (N_0 N_1)$.  In Eq. (\ref{eq:Evolution_chi})
a term of the order of $N^{-2}$ has been neglected, which is justified
for large values of $N$. The complete equation can be found in the
Supplementary Material \cite{SupportingInformation}.
Equation (\ref{eq:Evolution_chi}) is valid only for $ r > 0$.
The value for $r=0$ depends only on the fraction of black pixels $\phi$, $\chi(0) = \phi(1-\phi)$, and it is therefore a constant during the random walk.

Because each random move is independent from the previous one,
the analysis leading to Eq. (\ref{eq:Evolution_chi}) can be
repeated in an iterative way. Taking into account that the
starting two-point function is $\hat \chi (r)$, this leads to the following simple result
\begin{equation} \label{eq:chi_n_average}
\left< \chi (r) \right>^{(n)} = \hat  \chi(r) \ \alpha^n \ .
\end{equation}
In the course of the random walk, the average two-point
function therefore converges towards that of a Poisson point process
\cite{Serra:1982,Torquato:2000} with $ \chi(r) = 0$  for all $r >0$.
The convergence is exponential in the number of random moves and it
occurs in about $ N_0 N_1/(2 N)$ moves.

The determination of $\left<\chi^2(r)\right>^{(n)}$ proceeds along the same lines, but it is more involved; the details
can be found in the Supplementary Material \cite{SupportingInformation}. When the expression obtained for
$\left< \chi^2(r) \right>$ is introduced in Eq. (\ref{eq:E_n_average_general}), the value of the average energy
takes eventually the form
\begin{eqnarray} \label{eq:E_n_average}
\left< E \right>^{(n)}
= E_{\infty} + (E_2-E_1) \alpha^{n} + E_3 \beta^{n} \nonumber \\+ \left( E_1 - E_2 - E_3 - E_\infty \right) \gamma^n \ ,
\end{eqnarray}
where $E_1$, $E_2$, $E_3$ and $E_{\infty}$ are constants that characterize the starting ground state. The constants
$\beta$ and $\gamma$ depend only on $N$ and $N_1$
\begin{equation}
\beta = 1 - \frac{3 N}{N_1 N_0} \ ,
\quad
\gamma = 1 - \frac{4 N }{N_1 N_0} \left(1 - \frac{2}{N} \right)
\end{equation}
and $\alpha$ has the same meaning as in Eq. (\ref{eq:Evolution_chi}).

The constants $E_1$ and $E_{\infty}$ depend only on two-point characteristics of the ground states. They are written as
\begin{equation} \label{eq:E_constants}
E_1 = 2 \sum_{r} \hat \chi^2(r)
\end{equation}
and
\begin{equation} \label{eq:E_infty}
E_\infty = \sum_{r} \hat \chi^2(r) + \frac{2}{N} \frac{\phi^2(1-\phi)^2}{\omega_{r}} \ ,
\end{equation}
where the sum is over all the distances that are used in the definition of the energy. As a consequence
of Eq. (\ref{eq:E_n_average}), $E_\infty$ is the value towards which $\left< E \right>^{(n)}$ converges for
large values of $n$. Since the random walk is ergodic, any configuration has the same probability of being visited
in the long run. Therefore, $E_\infty$ is the average energy calculated over the entire configuration space,
which we refer to hereafter simply as $\left< E \right>$.

The main contribution to $\left< E \right>$ is $\sum \hat \chi^2$, which is small for disordered microstructures.
This term vanishes in the extreme case of a Poisson point process for which the only contribution left is of order
$1/N$, according to Eq. (\ref{eq:E_infty}). The shifting of the average energy towards lower values with increasing
disorder is clear in Figs. \ref{fig:DOS_13points} and \ref{fig:Random_Walk}.

The other two constants, $E_2$ and $E_3$, in the expression of $\left< E \right>^{(n)}$ depend on more than the
two-point function of the ground state. They are given by
\begin{eqnarray} \label{eq:E2}
E_2 = \frac{2}{N} \sum_{r}
2\phi(1 &-& \phi) \left\{\sigma^2(r) -\frac{\phi(1-\phi)}{\omega_{r}} \right\} \nonumber \\
&+& \frac{(1-2\phi)^2}{\omega_{r}} \ \hat \chi(r)
\end{eqnarray}
and
\begin{eqnarray} \label{eq:E3}
E_3 = \frac{4}{N} \sum_{r}
(1&-&2\phi)\phi \left\{
\sigma^2_C(r) - \sigma^2(r) + \frac{1}{\phi^2} \hat \chi^2(r) \right\} \nonumber \\
&-& \frac{(1-2\phi)^2}{\omega_{r}} \ \hat \chi(r) \ ,
\end{eqnarray}
where the functions $\sigma^2(r)$ and $\sigma_C^2(r)$ are defined as
\begin{equation} \label{eq:definition_sigma}
\sigma^2 (r) = \frac{1}{N} \sum_s \left\{
\frac{1}{\omega_r} \sum_i I(i) D_r(i,s) \right\}^2 - \phi^2
\end{equation}
and
\begin{equation} \label{eq:definition_sigma_C}
\sigma^2_{C} (r) = \frac{1}{N_1} \sum_s  I(s) \left\{
\frac{1}{\omega_r} \sum_i I(i) D_r(i,s) \right\}^2
- \left( \frac{S_2(r)}{\phi} \right)^2 \ .
\end{equation}
We postpone to Sec. \ref{sec:Roughness_metric} the detailed discussion of the structural meaning of $\sigma^2(r)$
and $\sigma^2_C(r)$. We should only mention here that $\sigma^2(r)$ can be expressed in terms of $S_2(r)$ and that
it is therefore common to all ground states (see the Supplementary Material \cite{SupportingInformation}).
By contrast, $\sigma^2_C(r)$ depends on three-point statistics and may differ from one ground state to another.

The black lines in Fig. \ref{fig:Random_Walk} have been obtained from Eq. (\ref{eq:E_n_average}) with the constants
$E_1$ to $E_\infty$ evaluated at the starting ground state. The analysis captures the essential features of the
random walk, in particular the steepness of the $\left< E \right>^{(n)}$ {\it versus} $n$ curves.

An important quantity for the rest of the analysis is the average energy of all configurations at Hamming
distance $d = 1$ from the ground state. Setting $n=1$ in Eq. (\ref{eq:E_n_average}) and neglecting terms of
the order of $N^{-3}$, leads to
\begin{equation} \label{eq:E1}
\left< E \right>^{(1)} = \frac{4 N}{N_1^2 N_0}
\sum_{r}
\Bigg\{ \hat \chi^2(r) + \phi^2 \ \sigma^2 (r)+ (1-2\phi) \phi^2 \ \sigma^2_{C} (r)
\Bigg\} \ .
\end{equation}
The first two contributions are global characteristics of the configuration space, which depend only on $S_2(r)$
and are therefore common to all ground states. By contrast, the contribution from $\sigma^2_C(r)$ may {\it a priori}
differ significantly from one ground state to another. We discuss this point in detail in Sec. \ref{sec:Roughness_metric}.

\subsection{The Energy Profiles of Individual Basins}
\label{sec:distances}

The average energies $\left< E \right>^{(n)}$ visited after exactly $n$ random moves is strictly a property of the
random walk, not only of the energy landscape. The aim of the present section is to use Eq. (\ref{eq:E_n_average})
to calculate the average energy of all microstructures at a given Hamming distance $d$ from a given ground state.

A random walk of length $n$ can reach any microstructure at Hamming distance $d \leq n$ from the ground state.
Let us call a $d$-state a microstructure at distance $d$ from the ground state, and $\nu_n(d)$ the fraction of all
the random walks of length $n$ that end on a $d$-state. The average energy $E(d)$ of all $d$-states is related to
$\left<E \right>^{(n)}$ through
\begin{equation} \label{eq:E_n_to_d}
\left< E \right>^{(n)} =  \sum_{d=0}^n \nu_n(d) E(d) \ .
\end{equation}
If the values of $\nu_n(d)$ were known, this relation could in principle be inverted to estimate $E(d)$ starting
from Eq. (\ref{eq:E_n_average}).

\begin{figure}
\begin{center}
\includegraphics[width=7cm, keepaspectratio]{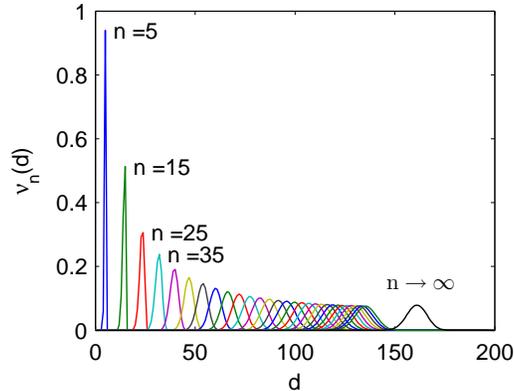}
\end{center}
\caption{(Color online) Distribution of Hamming distances to the ground state $\nu_n(d)$ for increasing number $n$
of steps in the random walk. The figure is for $N=1024$ and $N_1=200$, relevant to Fig. \ref{fig:Random_Walk}.}
\label{fig:RW_distance}
\end{figure}

The distribution $\nu_n(d)$ can be obtained by noticing that the random walk is a Markov process, and by calculating
the transition probabilities between various $d$-states. In real space, the Hamming distance is the minimum number
 of pixel displacements that is necessary to make the state identical to the ground state. Therefore, in a $i$-state,
 the black phase is identical to the ground state but for $i$ {\em holes}, and the white phase is identical to the
ground state but for $i$ {\em extra pixels}. Starting from an $i$-state at step $n$, there are $i^2$ ways to reach
 a $(i-1)$-state at step $n+1$. These correspond to the number of different ways to take one of the $i$ extra
pixels and place it into one of the $i$ holes. The transition probability is therefore
\begin{equation}
p_{ i \to (i-1) }=  i^2/(N_0 N_1) \ ,
\end{equation}
where the denominator is the total number of possible moves. There are $i (N-2i)$ different ways to reach
another $i$-state. These correspond to the $i(N_1-i)$ different ways of moving the holes within the black phase,
plus the $i(N_0-i)$ different ways of moving the extra pixels within the white phase. The transition probability
is therefore
\begin{equation}
p_{i \to i} = i(N-2i)/(N_0 N_1) \ .
\end{equation}
Finally, there are $(N_0-i)(N_1-i)$ different ways of reaching a $(i+1)$-state, which correspond to the different
ways of taking a black pixel and putting it in the white phase. This leads to
\begin{equation}
p_{i \to (i+1)}= (N_0-i)(N_1-i)/ (N_0 N_1) \ .
\end{equation}
The three transition probabilities add up to $1$, which proves that all possibilities have been considered.

The enumerated probabilities define a tridiagonal transition matrix $\mathbf{A}$ of size $(N_1+1)\times (N_1+1)$,
with elements $A(i,j) = p_{j \to i}$. The explicit form of $\mathbf A$ is given in the Supplementary Material
\cite{SupportingInformation}. Writing the values $\nu_n(d)$ in a vector as
$\underline{\nu}_{\ n} = \left[\nu_n(0),  \nu_n(1), \hdots ,\nu_n(N_1) \right]^T$ enables us to write
$ \underline{\nu}_{\ n+1} = \mathbf{A} \ \underline{\nu}_{\ n} $.
The general solution is therefore
\begin{equation}
\underline{\nu}_{\ n}= \mathbf{A}^n \underline{\nu}_{\ 0}
\end{equation}
where $\underline{\nu}_{\ 0}=\left[1 ,\ 0 ,\hdots , 0 \right]^T$ is the trivial distribution of Hamming
distances in the ground state.

The particular evolution of $\nu_n(d)$ for $N=1024$ and $N_1 = 200$ is shown in Fig. \ref{fig:RW_distance}.
These values are relevant to Fig. \ref{fig:Random_Walk}. For small values of $n$, the distribution is centered on
the value $d = n$. For large values of $n$, however, $\nu_n(d)$ converges towards an equilibrium distribution.
It is useful to note that although all states are accessible to the random walk after $n = N_1$ moves, the energy
$\left< E \right>^{(n)}$ continues to changes for larger values of $n$ (see Fig. \ref{fig:Random_Walk}) because
the convergence of $\nu_n(d)$ is asymptotic.

\begin{figure*}
\begin{center}
\includegraphics[width=12cm, keepaspectratio]{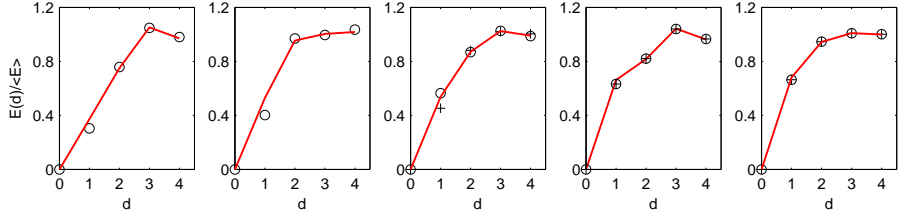} \\
\end{center}
\caption{(Color online) Energy profiles of the microstructures shown in Fig. \ref{fig:Degenerate}. From left to
right: $A$, $B$, $C_1$ ($\circ$) and $C_2$ ($+$), $D_1$ ($\circ$) and $D_2$ ($+$), $E_1$ ($\circ$) and $E_2$ ($+$).
 Note that the profiles of $D_1$ and $D_2$, as well as $E_1$ and $E_2$ are identical. The solid line is the approximate
profile, common to all ground states, calculated from $\hat S_2(r)$ alone using Eq. (\ref{eq:sigma2C_approximation}).}
\label{fig:Profiles_N4}
\end{figure*}

Using the known values of $\nu_n(d)$ and of $\left<E \right>^{(n)}$, Eq. (\ref{eq:E_n_to_d}) can be inverted for
$n = 1, 2, ..., N_1$, yielding the values of $ E (d)$. The procedure is illustrated in Fig. \ref{fig:Profiles_N4}
 for the small-system-size configurations of Fig. \ref{fig:Degenerate}. The results are plotted in the normalized
form $E(d) / \left< E \right>$, which we refer to as the energy profiles.

The energy profiles describe quantitatively the average energy landscape surrounding any particular ground state. They
are all initially increasing curves that start from $0$ and reach values close to $1$. The average energy for $d=1$
is equal to $\left< E \right>^{(1)}$, as calculated from Eq. (\ref{eq:E1}). For large Hamming distances the energy
decreases again because microstructures with large $d$ can be thought of as negative imprints of the ground state:
for $d = N_1$ the points that were occupied by black pixels in the ground state are all occupied by white pixels.

\begin{figure*}
\begin{center}
\includegraphics[width=12cm, keepaspectratio]{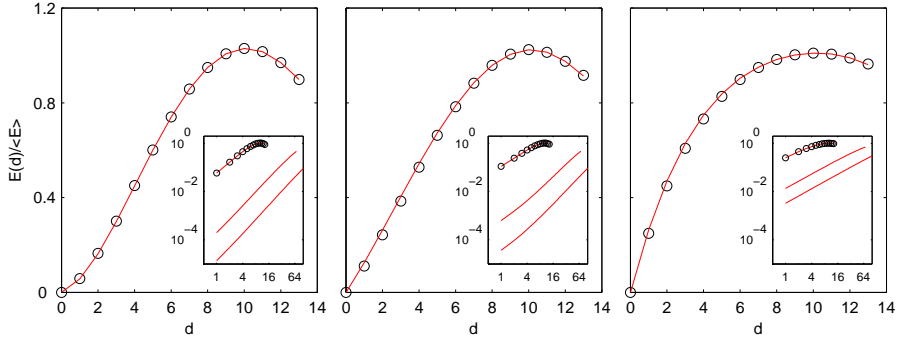} \\
\end{center}
\caption{(Color online.) From left to right: energy profiles of the single disk, hard-disks, and Poisson microstructures
shown in Fig. \ref{fig:DOS_13points}. The symbols ($\circ$) are the exact values and the solid lines are the approximate
profile, common to all ground states, calculated from $\hat S_2(r)$ alone using Eq. (\ref{eq:sigma2C_approximation}). The exact values are plotted
in insets on logarithmic scale, together with the profiles of equivalent systems of larger sizes, namely with $L =32$ (top)
and $L=64$ (bottom). The microstructures used for the $L=32$ profiles are those of Fig. \ref{fig:REC_examples}.
The approximate and exact profiles are indistinguishable on the scale of the insets.}
\label{fig:Profiles_N13}
\end{figure*}

Figure \ref{fig:Profiles_N13} shows the energy profiles of the single disk, the hard disks, and the Poisson point
process of Fig. \ref{fig:DOS_13points}. When plotted on logarithmic scales (insets), the profiles are seen to satisfy
a power law of the type
\begin{equation}
E(d) = E(1) d^\delta
\end{equation}
for small values of $d$. When the resolution is increased - i.e. increasing $N$ from $8^2$ to $N =32^2$ and $N=64^2$ while
keeping $N_1/N$ constant - the profiles are shifted vertically (insets) but the exponent of the power law persists.
In the case of the reconstruction of the single disk, the exponent $\delta$ is close to $2$ ($\approx 1.97$), and for
the reconstruction of the Poisson point process the exponent is close to $1$ ($\approx 0.92$). The energy profile of the
hard disks is not a pure power law: the exponent is that of a single disk for large Hamming distances and that of a
Poisson point process for shorter distances.

The different exponents $\delta$ for the Poisson point process and for the single disk hint at a qualitative difference
that can be understood with the {\em hole} and {\em extra pixel} interpretation of Hamming distance $d$. In the case of
the Poisson point process, the energy is proportional to $d$. This means that any hole added to the ground state contributes
additively to the energy, which points to the absence of effective pixel-pixel interaction energy. By contrast, the
quadratic behavior for the single disk points to a collective contribution of the pixels to the overall energy, which
can be considered as the signature of a structure.

\section{Energy Roughness as a Proxy for Ground-State Degeneracy}
\label{sec:Roughness_metric}

When comparing the energy profiles in Figs. \ref{fig:Profiles_N4} and \ref{fig:Profiles_N13} with the densities of states
in Figs. \ref{fig:DOS_4points} and \ref{fig:DOS_13points}, a striking correlation appears between the sizes of the basins
and the ground-state degeneracy $\Omega_0$. We observe that the smaller the basin, the more degenerate the reconstruction.
This observation is consistent with the one that large ground-state degeneracies are generally associated with rough
energy landscapes \cite{Wales:1998,Debenedetti:2001}.

However, a major difference between the energy profiles and the ground-state degeneracy is that the latter is a global
characteristic of the reconstruction problem but the former are specific to given ground states. For example, in the case
of configuration $C$ of Fig. \ref{fig:Degenerate}, the ground states $C_1$ (the Kite) and $C_2$ (the Trapezoid) have
slightly different energy profiles (Fig. \ref{fig:Profiles_N4}). The main purpose of the present section is
to provide an approximation for the energy profile, common to all ground states, which can be calculated from $\hat S_2(r)$ alone.

To understand how the energy profiles depend on the particular ground state, it is necessary to analyze the structural meaning of functions $\sigma^2(r)$ and $\sigma^2_C(r)$ defined by Eqs. (\ref{eq:definition_sigma}) and (\ref{eq:definition_sigma_C}). We show in the Supplementary Material \cite{SupportingInformation} that $\sigma^2(r)$ can be expressed in terms of the
two-point function $\chi(r)$ as
\begin{equation} \label{eq:sigma2_chi}
\sigma^2 (r) = \sum_{l \leq 2r} v(r,l) \omega_l \chi(l) \ ,
\end{equation}
where $v(r,l)$ is a characteristic of the grid and of the boundary conditions. By definition, all ground states have
identical two-point statistics. The contribution of $\sigma^2(r)$ to the energy profile is therefore common to all ground
states. By contrast, it results from Eq. (\ref{eq:definition_sigma_C}) that $\sigma^2_C(r)$ is a sum of terms of the type
\begin{equation} \label{eq:triangle}
I(s) I(i) I(j) D_r(s,i) D_r(s,j) \ ,
\end{equation}
which incorporate three-point statistics. Accordingly, the contribution of $\sigma^2_C(r)$ to the energy profile may
differ significantly from one ground state to another.

As a consequence of Eq. (\ref{eq:triangle}), the pixel configurations that contribute to $\sigma^2_C(r)$ are isosceles
triangles with apex $s$ and side-length $r$. In the case of configuration $C$ of Fig. \ref{fig:Degenerate}, the Kite
($C_1$) has two such triangles, one with $r = 1$ and the other with $r = \sqrt{5}$. On the contrary there is no isosceles
triangle in the Trapezoid ($C_2$). It therefore results from Eq. (\ref{eq:E1}) that $\left<E\right>^{(1)}$ is larger
for $C_1$ than for $C_2$, in agreement with Fig. \ref{fig:Profiles_N4}.

The energy profiles of $D_1$ and $D_2$ are identical although these two ground states are different. This originates
 in the fact that in both ground states, there is an isosceles triangle of side $r = 2$ and another one of
side $r = \sqrt{10}$. The same explanation applies to $E_1$ and $E_2$. In both ground states there is a single
isosceles triangle with $r = \sqrt{5}$.

Despite these differences between $\sigma^2(r)$ and $\sigma^2_C(r)$, the two functions have a strong similarity
which can be put in evidence by the following probabilistic interpretation. Consider the set of all pixels at distance
 $r$ from a given pixel $s$, which we refer to as the ring centered on $s$. The fraction of black pixels in the ring
can be written as
\begin{equation} \label{eq:varphi}
\varphi_r = \frac{1}{\omega_r} \sum_{i=1}^N I(i) D_r(i,s) \ .
\end{equation}
When $s$ is chosen randomly among all pixels (black and white) $\varphi_r$ is a random variable having average value
$\phi$. Equation (\ref{eq:definition_sigma}) shows that $\sigma^2(r)$ is the variance of $\varphi_r$. The function
$\sigma^2(r)$ can therefore be thought of as a generalized coarseness \cite{Lu:1990}.

From this probabilistic perspective, the meaning of $\sigma^2_C(r)$ is equivalent to $\sigma^2(r)$, only the central
pixel $s$ is not distributed randomly over the entire space but only over the black pixels. In this case, the average of
$\varphi_r$ is the conditional probability that a pixel of the ring is black, given that the central pixel is black too,
i.e. $S_2(r)/\phi$. From Eq. (\ref{eq:definition_sigma_C}), one sees that $\sigma^2_C(r)$ is the variance of $\varphi_r$
when the central pixel $s$ is randomly distributed over the black phase. The function $\sigma^2_C(r)$ can therefore be
thought of as a conditional coarseness.

\begin{figure}
\begin{center}
\includegraphics[width=7cm, keepaspectratio]{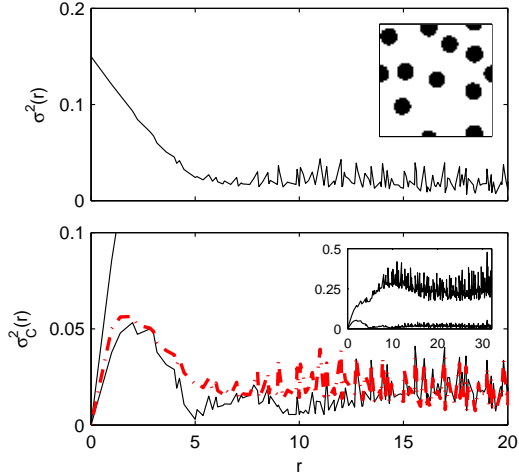} \\
\end{center}
\caption{(Color online) Functions $\sigma^2(r)$ (top) and $\sigma^2_C(r)$ (bottom) of a hard-disk microstructure. The
inset of the bottom graph shows $\sigma^2_C(r)$ and its upper bound $\sigma^2_U(r)$. The dashed red line is the
approximation $\tilde \sigma^2_C(r)$ obtained through Eq. (\ref{eq:sigma2C_approximation}).}
\label{fig:Sigma2}
\end{figure}

For small radii $r$, the values taken by $I(i)$ on the ring are highly correlated with the value in the center, which
implies $\sigma^2_C(r \to 0) = 0$. For large radii, the values on the ring and in the center are not correlated at all
and therefore $\sigma^2_C(r) \simeq \sigma^2(r)$. An example of $\sigma^2(r)$ and $\sigma_C^2(r)$, calculated on the
realization of a hard-disk system is given in Fig. \ref{fig:Sigma2}.

The similarity of the probabilistic interpretations of $\sigma^2_C(r)$ and of $\sigma^2(r)$, and their strict mathematical
equality for large values of $r$, suggest that it should be possible to find an approximation of $\sigma^2_C(r)$ in terms
of two-point functions only. This would enable us to estimate a single approximate energy profile that would depend only
on $\hat S_2(r)$. That profile would therefore be common to all ground states of a given reconstruction problem.

To find such an approximation, we observe that the terms between braces in the definitions of $\sigma^2_{C}(r)$  and
$\sigma^2(r)$ are identical and that they are all positive (see Eqs (\ref{eq:definition_sigma}) and
(\ref{eq:definition_sigma_C})). However, there are fewer terms in Eq. (\ref{eq:definition_sigma_C}) because $I(s)$
can be equal to $0$. One has therefore
\begin{equation}
N_1 \left(
\sigma^2_{C} (r) + \frac{S_2^2(r)}{\phi^2}
\right)
\le
N \left(
\sigma^2 (r) + \phi^2
\right) \ ,
\end{equation}
which leads to the following upper bound for $\sigma^2_C(r)$:
\begin{equation} \label{eq:sigma2C_bounds}
\sigma^2_C(r) \leq \sigma^2_{U} (r) = \frac{1}{\phi} \sigma^2 (r) + \phi - \frac{S_2^2(r)}{\phi^2} \ ,
\end{equation}
which depends only on two-point statistics.

The inset of Fig. \ref{fig:Sigma2} compares $\sigma^2_C(r)$ to $\sigma^2_U(r)$ in the particular case of a hard-disk
microstructure. The upper bound $\sigma^2_U(r)$ is a good approximation of $\sigma^2_C(r)$ only for very small $r$.
We therefore propose the following approximation for $\sigma^2_C(r)$
\begin{equation} \label{eq:sigma2C_approximation}
\tilde \sigma^2_C(r) = \left(\frac{1}{\sigma^2_U(r)} + \frac{1}{\sigma^2(r)} \right)^{-1} \ ,
\end{equation}
which is practically equal to $\sigma^2(r)$ for large $r$, as it should. Fig. \ref{fig:Sigma2} shows that
$\tilde \sigma^2_C(r)$ is a fair approximation of $\sigma^2_C(r)$ at all radii.

Using $\tilde \sigma^2_C(r)$ in place of $\sigma^2_C(r)$ enables us to calculate a single energy profile, based on
$\hat S_2(r)$ alone. The red curves in Figs. \ref{fig:Profiles_N4} and \ref{fig:Profiles_N13} have been calculated
in that way: $\sigma^2(r)$ was calculated rigorously from the target $\hat S_2(r)$ through Eq. (\ref{eq:sigma2_chi}),
and $\sigma_C^2(r)$ was approximated by Eq. (\ref{eq:sigma2C_approximation}). In the case of the larger microstructure
shown in the insets of Fig. \ref{fig:Profiles_N13}, the approximate profiles are indistinguishable from the exact
profiles on the scale of the figure.

\begin{figure}
\begin{center}
\includegraphics[width=7cm, keepaspectratio]{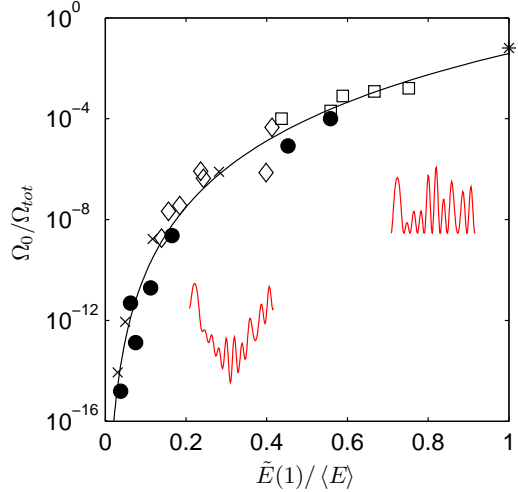} \\
\end{center}
\caption{(Color online) Relation between ground-state degeneracy $\Omega_0/\Omega_{tot}$ and roughness of the energy landscape $\tilde E(1)/\left< E \right>$ calculated from $S_2(r)$ alone. The various microstructures are: disks of different sizes ($\bullet$),
realizations of Poisson point processes of various densities ($\diamond$), hard-disk microstructures ($\times$),
the configurations of Fig. \ref{fig:Degenerate} with $N_1=4$ ($\square$), as well as a configuration with $N_1 =2$ ($*$).
The black line is a guide to the eye; the insets are sketches of archetypical energy landscapes for large and small
values of $\tilde E(1)/\left< E \right>$.}
\label{fig:Roughness}
\end{figure}

Using Eq. (\ref{eq:sigma2C_approximation}) it is possible to estimate a single metric to characterize globally the
roughness of the energy landscape. We propose the ratio
\begin{equation} \label{eq:roughness_metric}
\tilde E(1) / \left< E \right> \ ,
\end{equation}
where the tilde highlights the fact that $\tilde E(1)$ is estimated through the approximation $\tilde \sigma^2_C(r)$.
The quantity $\tilde E(1)$ is the average energy of all microstructures at distance $d =1$ from the ground state. Because
the ground states have zero energy, $\tilde E(1)$ can be thought of as a Laplacian in configuration space $\mathcal{C}$.
The ratio $ \tilde E(1)/\left< E\right>$ is therefore a dimensionless measure of the total curvature of the energy
surface in the vicinity of any ground state. It has to be stressed that $\tilde E(1) / \left< E \right>$ is calculated from $S_2(r)$ alone, and that it is therefore not specific to any particular ground state.

Figure \ref{fig:Roughness} shows the quantitative relation between $\tilde E(1)/\left< E\right>$ and the normalized
ground-state degeneracy $\Omega_0/\Omega_{tot}$ for a variety of microstructures defined on a $8 \times 8$ grid. The
microstructures used for the figure are available in the Supplementary Material \cite{SupportingInformation}: they
comprise both non-degenerate disk-like objects and highly degenerate realizations of Poisson point processes.
The ground-state degeneracy of the latter was estimated via the MC algorithm. The quantity $\Omega_0/\Omega_{tot}$
is found to be highly correlated with $\tilde E(1)/E_\infty$ over more than 14 orders of magnitude.

When passing from small to large values of $\tilde E(1)/\left< E \right>$, the energy landscape changes qualitatively
in the way suggested by the insets of Fig. \ref{fig:Roughness}. For low values of $\tilde E(1)/\left< E \right>$,
the energy landscape has an overall funnel structure, with low energy barriers, which makes it well suited for
 optimization problems. By contrast, for large values of $\tilde E(1)/\left< E \right>$, the landscape is very rough
with a large ground-state degeneracy. It is, however, interesting to note that the rightmost point in Fig. \ref{fig:Roughness}
is obtained for a system with $N_1 = 2$ having thus only a trivial degeneracy. The corresponding energy landscape
is extremely rough because any possible energy can be found at a Hamming distance as short as $d = 1$ from the ground
state, but the total number of configurations $\Omega_{tot}$ is also extremely small.

The data referred to as disks in Fig. \ref{fig:Roughness} is a collection of non-degenerate microstructures with increasing
values of $N_1$. When increasing $N_1$, the roughness $\tilde E(1)/\left< E \right>$ decreases but the degeneracy remains
equal to its trivial translation contribution $\Omega_0 = N$. It is noteworthy that the values of $\Omega_0/\Omega_{tot}$
of these non-degenerate microstructures span the same curve as the realizations of Poisson point processes, for which
$\Omega_0$ has a huge non-trivial contribution.

\begin{figure}
\begin{center}
\includegraphics[width=7cm, keepaspectratio]{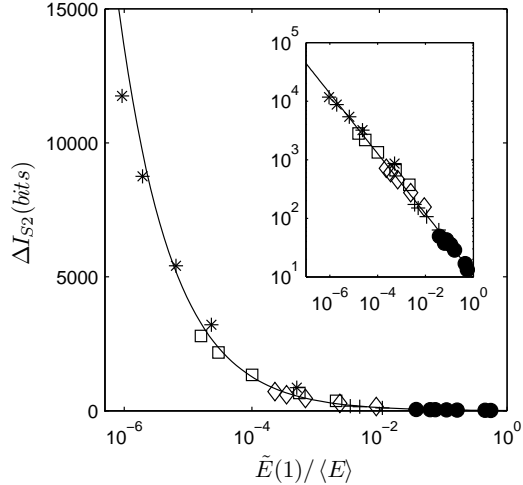} \\
\end{center}
\caption{Relation between the amount of structural information $\Delta I_{S2}$ in the two-point function and the roughness
metric $\tilde E(1)/\left< E \right>$. The relation was obtained from disks of increasing sizes $N_1$ defined on grids of
size $N = 8 \times 8$ ($\bullet$), $16\times 16$ ($+$), $32\times 32$ ($\diamond$),
$64\times 64$ ($\square$) and $128 \times 128$ ($*$). The inset shows the data on logarithmic scales. The solid line is
 Eq. (\ref{eq:Delta_I_S2_fit}) obtained by least-square fit.}
\label{fig:Information}
\end{figure}

The microstructures considered in Fig. \ref{fig:Roughness} were limited to $8 \times 8$ grids, which size limit is
imposed by the convergence of the MC algorithm. However, the fact that the
$\Omega_0/\Omega_{tot}$-{\it versus}-$\tilde E(1)/\left< E \right>$ curve does not discriminate trivial from non-trivial
degeneracy should not be limited to small microstructures. This assumption enables us to extend the curve to larger
microstructures by using disks of increasing sizes $N_1$, on grids with increasing sizes $N$, for which the degeneracy
is known to be exactly $\Omega_0 = N$. This was done in Fig. \ref{fig:Information}. The degeneracy is plotted in the form
of $\Delta I_{S2} = \log_2(\Omega_{tot}/\Omega_0)$ for reasons that will be explained in next section. The inset of
the figure shows that disks of all sizes and on all grids span a single curve which obeys approximately a power law.

\section{Degeneracy Analysis Using an Information-Theoretic Formulation}

The degeneracy $\Omega_0$ can be analyzed in terms of the information content associated with a given two-point function
$S_2(r)$. Indeed, if a reconstruction problem is non-degenerate, the two-point correlation function is a complete
characterization of the microstructure. By contrast, in the case of a large degeneracy the correlation function contains
 a relatively small amount of microstructural information.

This idea can be made quantitative by borrowing concepts from information theory \cite{Cover:1991,Berut:2012}.
In that context, a given microstructure is considered to be the outcome of a random process. More specifically, if
nothing is known other than the total number of pixels $N$, then specifying a given microstructure is equivalent to drawing
it out of the complete configuration space. Any microstructure is therefore an event having probability $ p = 1 / 2^N$.
The self-information (or so-called {\it surprisal}) associated with such an event is
\begin{equation}
I = - \log_2(p) = N \ .
\end{equation}
The use of a base 2 logarithm ensures that the self-information is expressed in units of {\em bits}. The self-information can
be used as a quantitative measure for the information content of the realization of an event. In this particular case,
the value is $I =N$ bits, i.e., 1 bit per pixel, which is quite consistent.

\begin{table*}
\caption{\label{tab2} Information-theoretic analysis of the reconstructions of Figs. \ref{fig:REC_examples} and
\ref{fig:REC_crystal}, with $N$ and $N_1$: total number of pixels and number of black pixels; $\Delta \tilde I_{S1}$:
fractional information content of the one-point statistics; $\tilde E(1)/\left< E\right>$: roughness metric, calculated from $S_2(r)$ alone;
$\Delta \tilde I_{S2}$: fractional information content of the two-point statistics; $\Delta \tilde I_{S1} + \Delta \tilde I_{S2}$:
total information available for the reconstruction.}
\begin{ruledtabular}
\begin{tabular}{lccccccccc}
Microstructure & $N$ & $N_1$ & $\Delta \tilde I_{S1}$ & $\tilde E(1)/ \left< E \right>$  & $\Delta \tilde I _{S2}$ & $\Delta \tilde I _{S1}+\Delta \tilde I _{S2}$ \\
\hline
Single disk & 1024 & 200 & 0.29 & $2.27 \ 10^{-4}$  & 0.81 &  1.10 \\
Hard disks & 1024 & 200 & 0.29 & $ 6.38 \ 10^{-4}$  & 0.48 &  0.77 \\
Poisson point process & 1024 & 200 & 0.29 & $1.33 \ 10^{-2}$  & 0.10 &  0.39 \\
\hline
Crystal & 16384 & 3000 & 0.31 & $1.79 \ 10^{-6}$  & 0.61 &  0.93 \\
Polycrystal 1 & 16384 & 3000 & 0.31 & $ 6.33 \ 10^{-6}$  & 0.32 &  0.63 \\
Polycrystal 2 & 16384 & 3000 & 0.31 & $1.36 \ 10^{-5}$  & 0.22 &  0.53 \\
\end{tabular}
\end{ruledtabular}
\end{table*}

If the number of black pixels $N_1$ is known, a microstructure is no longer a draw out of $2^N$, but rather out of
$\Omega_{tot} = \binom{N}{N_1}$. This implies that the self-information is reduced by a quantity
\begin{equation} \label{eq:Delta_I_S1}
\Delta I_{S1} = N - \log_2\left( \Omega_{tot} \right) \ ,
\end{equation}
which can be thought of as the amount of information contained in specifying the value of $N_1$, compared to knowing merely
the system size. We refer to $\Delta I_{S1}$ as the information content of the one-point statistics. The quantity
$\Delta I_{S1}$ depends on the particular value of $N_1$. It is equal to $N$ for $N_1=0$ and for $N_1 = N$. In both
cases specifying $N_1$ is a complete description of the microstructure, since in those cases the system is either
completely black or completely white. By contrast, $\Delta I_{S1}$ is minimum for $N_1 = N/2$, because this particular
value maximizes $\Omega_{tot}$.

The same reasoning can be applied to quantify the information content of $S_2(r)$. Once $S_2(r)$ is given, a particular
microstructure is drawn out of $\Omega_0$ possibilities, and no longer $\Omega_{tot}$. This suggests defining the quantity
\begin{equation}  \label{eq:Delta_I_S2}
\Delta I_{S2} = \log_2\left( \Omega_{tot} \right) - \log_2\left( \Omega_0 \right)
\end{equation}
to measure the information content of $S_2(r)$, in addition to knowing $N_1$.

Note that the information content of the one-point and two-point statistics can be understood in terms
of configurational entropies corresponding to different definitions of the macrostate of the system. In the case of the
one-point information, the macrostate is specified via the value of $N_1$, which results in an entropy $\log_2(\Omega_{tot})$.
Similarly, if $S_2(r)$ is used to define the macrostate, the entropy becomes $\log_2(\Omega_0)$. The information content
$\Delta I_{S2}$ is equal to the entropy reduction that results from incorporating $S_2(r)$ in the definition
of the macrostate of the system.

The analysis of Sec. IV suggests that $\Delta I_{S2}$ can be accurately calculated from $\tilde E(1)/\left< E \right>$ alone.
This is the significance of Fig. \ref{fig:Information}. The inset of the figure shows that the dependency is a power-law
of the type
\begin{equation}  \label{eq:Delta_I_S2_fit}
\Delta I_{S2} {\textrm [bits]} \simeq 11.2 \left( \frac{\tilde E(1)}{\left< E \right>} \right)^{-0.51} \ ,
\end{equation}
where the numerical coefficients have been obtained by a least-square fit. It has to be stressed here that
the roughness metric $\tilde E(1)/\left< E \right>$ is calculated from $S_2(r)$ alone. Therefore, Eq. (\ref{eq:Delta_I_S2_fit}) provides a practical means to estimate $\Delta I_{S2}$ in any experimental context where the only information about the system is its correlation function $S_2(r)$.

For a reconstruction to be accurate, the total information available in the form of one-point and two-point statistics,
has to be $N$ bits. Therefore, the utility of $\Delta I_{S1}$ and $\Delta I_{S2}$ is best illustrated by normalizing them
by $N$ and defining the following fractional information contents $\Delta \tilde I_{S1} = \Delta I_{S1}/N $ and
$\Delta \tilde I_{S2} = \Delta I_{S2}/N $, which take values between 0 and 1. A reconstruction is accurate whenever the sum
$\Delta \tilde I_{S1} + \Delta \tilde I_{S2}$ is close to one. Table \ref{tab2} analyzes the reconstruction examples of
Figs. \ref{fig:REC_examples} and \ref{fig:REC_crystal} along these directions.

The three microstructures of Fig. \ref{fig:REC_examples} all have $N_1 = 200$ black pixels on a grid with a total of
$N = 1024$ pixels. In this case, one estimates through Eq. (\ref{eq:Delta_I_S1}) that the one-point information is
$\Delta \tilde I_{S1} \simeq 0.29$. The two-point information $\Delta I_{S2}$ for the various microstructures was calculated
from the roughness metric $\tilde E(1) / \left< E \right>$ through Eq. (\ref{eq:Delta_I_S2_fit}) and the corresponding
values of $\Delta \tilde I_{S2}$ are reported in Table II. In the case of the single disk, the total information content
of $S_1$ and $S_2$ is close to 1, which means that a perfect reconstruction is possible. The fact that the value is
slightly larger than 1 results from the limited accuracy of Eq. (\ref{eq:Delta_I_S2_fit}). In the case of the Poisson point
process, the total information available amounts to only 39\% of the information required for the reconstruction: the
reconstruction is therefore impossible.

The case of the hard disks is not so clear-cut: the reconstruction captures many structural characteristics of the target
(Fig. \ref{fig:REC_examples}) although only 77\% of the information is available (Table \ref{tab2}). This seems to suggest
that a fair reconstruction may be possible with about 20\% of missing structural information.

The information-theoretic analysis of the polycrystal reconstructions of Fig. \ref{fig:REC_crystal} proceeds in the same way.
In the case of the single crystal, 93\% of the information is (see Table \ref{tab2}), which is consistent with the good
quality of the reconstruction. For decreasing crystallite sizes, the amount of information decreases. In the case of the
smallest crystallites, the amount of information is only about 50 \% and the reconstruction is expectedly inaccurate.

The consistency of the information analysis of the single disk reconstruction was expected because Fig. \ref{fig:Information}
and Eq. (\ref{eq:Delta_I_S2_fit}) are based on disks of various sizes. The validity of Eq. (\ref{eq:Delta_I_S2_fit}) for
microstructures other than disks is established only for very small systems, for which the MC algorithm could be used
(Fig. \ref{fig:Roughness}). The fact the present analysis enables us to predict the non-degeneracy of the crystal
reconstruction strongly supports the generality of Eq. (\ref{eq:Delta_I_S2_fit}).

\section{Discussion and Conclusions}

Throughout the paper, we have discussed several cases of trivially and non-trivially degenerate microstructures. We have
argued that the geometrical features that contribute to decreasing the non-trivial contribution to $\Omega_0$ are those
that lead to extremal values of $S_2(r)$ for given values of $r$. This is notably the case for a single disk under periodic
boundary conditions, which is the microstructure that maximizes $S_2(r)$ for sufficiently small $r$'s. The crystal
configuration shown in Fig. \ref{fig:REC_crystal} is non-degenerate for similar reasons: because of its anisotropy,
that microstructure achieves extremal values of $S_2(r)$ for many different values of $r$. The opposite situation is
that of a Poisson point processes, for which $S_2(r)$ takes values close to the average value $\phi^2$ for all $r>0$.
This leads to a huge degeneracy (configurational entropy) because none of the values of $S_2(r)$ is close to being extremal.

It is interesting to note that the single disk within a periodic box can be thought of as a dilute distribution of
disks in all of Euclidean space when the infinite number of periodically replicated cells are considered. In the case of
impenetrable disks at arbitrary density, it is noteworthy that $S_2(r)$ can be exactly expressed in terms of a one-body
or low-density term (which contains the same shape and surface area information as in the dilute regime) and a single
higher-order two-body term involving pair statistics \cite{Torquato:2000,Torquato:1985}. It is therefore the latter
term that is responsible for the degeneracy of such configurations for arbitrary densities.

The trivial contribution to the degeneracy $\Omega_0$ depends on the particular rotational symmetry and chirality of
the microstructure, but it is always of the order of total number of pixels in the grid $N$. By contrast, the
non-trivial contribution to the degeneracy can be significantly larger. The Monte Carlo estimation of $\Omega_0$
shows that even a  modestly sized $8 \times 8$ Poisson point process can have a degeneracy as large as
$\Omega_0 \approx 10^7$ (see Fig. \ref{fig:DOS_13points}). This value is expected to increase exponentially
with the size of the system because any possible $S_2$-preserving pixel displacement contributes
multiplicatively to $\Omega_0$.

In order to quantitatively address the question of the degeneracy corresponding to any specified correlation function,
we have mapped it to the determination of a ground-state degeneracy. This mapping led us to two breakthroughs.
First, we now can  calculate for the first time the density of states for reconstruction problems via a Monte Carlo
algorithm, and in particular to determine the values of $\Omega_0$ for a few benchmark systems. Second, we built on
the general observation throughout physics that large ground-state degeneracies are generally associated with rough
energy landscapes, which enabled us to use the roughness of the energy landscape as a proxy for the microstructural degeneracy.

A natural metric for the roughness of the energy landscape is the total curvature of the energy surface, evaluated at
the ground-states. Using a random walk in configuration space (see Fig. \ref{fig:Random_Walk}), we derived an analytic
expression for the total energy-surface curvature in the form of $\tilde E(1)/\left< E \right>$, which can be calculated
in terms of $S_2(r)$ alone. The Monte Carlo analysis confirms that $\tilde E(1)/\left< E \right>$ is indeed highly
correlated with the degeneracy of a reconstruction problem, independently of the type or microstructure considered
(Fig. \ref{fig:Roughness}). It has to be noted that the roughness metric is consistent with the intuitive analysis of
degeneracy in terms of extremal values of $S_2(r)$. Indeed, the main contribution to the denominator $\left< E \right>$
is $\sum \chi^2(r)$, so that any value of $S_2(r)$ larger or smaller than $\phi^2$ contributes to decreasing the
roughness metric, and hence the degeneracy.

A counterintuitive result of the present study is that the distinction between trivial and non-trivial degeneracy
is irrelevant in configuration space $\mathcal{C}$. In particular, the quantitative relationship found between $\Omega_0$
and the roughness metric does not discriminate the two types of degeneracy. This enabled us to use trivially-degenerate
microstructures to generate a single calibration relation for $\Omega_0$ as a function of $\tilde E(1)/\left< E \right>$.
That relation applies to a large variety of microstructures of sizes much larger than those analyzable by the Monte
Carlo method (see Fig. \ref{fig:Information}).

We should point out that although the examples discussed in the present work are all two-dimensional microstructures, the same methodology can be applied in any space dimension.
It is indeed noteworthy that Eq. (\ref{eq:E1}) and the approximation Eq. (\ref{eq:sigma2C_approximation}) are valid in any space dimension. As a consequence, the roughness metric of any higher-dimensional microstructure can be calculated easily from its correlation function $S_2(r)$ alone. Moreover, the observation that we make that the relation between the roughness metric and the ground-state degeneracy does not discriminate trivial from non-trivial degeneracy is also expected to hold in any space dimension. Therefore, higher-dimensional trivially-degenerate microstructures (e.g. hyperspheres) can be used to produce a relation equivalent to Eq. (\ref{eq:Delta_I_S2_fit}) or Fig. \ref{fig:Information} in any space dimension.

It is also noteworthy that our analytical results do not assume Euclidean space: the only restriction is that $\sum_j D_r(i,j)$ should be independent of $i$, where $D_r(i,j)$
is the operator used to define $S_2(r)$ through Eq. (\ref{eq:definition_P}). Therefore, the mathematical expression of
the roughness metric is valid in hyperbolic and spherical spaces as well as in any dimension. However, the relationship
 between the degeneracy (configurational entropy) and the roughness metric is expected to be space- and dimension-dependent.

The use of information-theoretic concepts allows our methods to be easily
applied in practice. As mentioned in the introduction, two-point correlation functions are often the only data
available experimentally for in situ studies with a nanometer resolution, notably through small-angle
scattering measurements \cite{Svergun:1999, Svergun:2001,Beale:2006}. The question of the structural ambiguity
of small-angle scattering patterns is an old one \cite{Patterson:1939,Volkov:2003,Gommes:2008}, but the recent
development of very intense X-ray sources \cite{Bostedt:2012} has ignited a very lively debate about the
possibility of reconstructing nanometer-scale objects from scattering patterns \cite{Reich:2011}.
Our analysis provides a novel and very general approach to address this type of question: An accurate
reconstruction is possible whenever the amount of information $\Delta \tilde I_{S1} + \Delta \tilde I_{S2}$
 is close to one. The examples that we have discussed suggest that a relatively accurate reconstruction is
 possible with up to 20\% missing information, but it is premature to formulate any general rule.

It has to be stressed that, although the present work is based on the reconstruction of microstructures defined
on a grid with exact distances, the results apply unchanged to the discrete reconstruction of microstructures
starting from experimental (i.e., continuous) correlation functions. Correlation functions of the type of the
monocrystal (Fig. \ref{fig:REC_crystal}) are unrealistic in an experimental context. However, the general
relation between $\tilde E(1)/\left< E \right>$ and $\Delta I_{S2}$ still holds. The only difference is that
experimental correlation functions of disordered systems are generally of the polycrystal type, with very small
crystallites. Except in some exceptional cases, it is therefore expected that experimental correlation functions
 with no orientation information should be highly degenerate.

The domain of applications of our results is not limited to scattering. Other applications can notably be
found in the field of computer vision for texture recognition. A texture with low degeneracy $\Omega_0$ can
in principle be discriminated robustly based on two-point statistics alone, which would make slower
three-point characterizations unnecessary \cite{Yellott:1993}.

Besides applications, information-theoretic concepts are also useful conceptually. It is very natural that a reconstruction
be possible whenever the information content of the available data is equal to the amount of information required,
i.e. $N$ bits where $N$ is the total number of pixels. In the cases we considered, the information came in the
form of one-point statistics $\Delta I_{S1}$ and of two-point statistics $\Delta I_{S2}$. However, the approach
could be generalized naturally to higher-order statistics \cite{Matheron:1967,Torquato:1982,Torquato:1983,Torquato:1988B}.
Quite generally, a successful reconstruction would require all correlations to be considered up to the $m$th order,
with $m$ satisfying
\begin{equation} \label{eq:Information_Series}
\sum_{n=1}^m \Delta I_{Sn} \approx N
\end{equation}
where $\Delta I_{Sn}$ is the information contained in n-point correlation function $S_{n}$ in addition to $S_{n-1}$.

There is some evidence supporting the view that $S_3$ does not contain significant information in addition to $S_2$
 \cite{Aubert:2000,Jiao:2009}. In the present context this suggests that the series in Eq. (\ref{eq:Information_Series})
converges slowly. The approach could be further generalized to other types of statistical descriptors including lineal
statistics \cite{Mering:1968,Torquato:1993}, pore-size functions \cite{Serra:1982,Torquato:2000}, and cluster correlation
functions \cite{Torquato:1985,Jiao:2009}.

We stress that the present work has numerous ramifications in materials sciences and beyond. For instance, an important
question concerns the realizability of two-point correlation functions \cite{Torquato:2006,Quintanilla:2008}. It would be
interesting to explore whether new necessary conditions for the realizability of $S_2(r)$ can be obtained by expressing
that the information content (in bits) cannot exceed the total number of pixels in the microstructure.

Last but not least, other applications can be found in physics. Indeed, the Hamiltonian of any system with pairwise
additive energy can be written as
\begin{equation} \label{eq:pairwise_energy}
H = \sum_r v(r) S_2(r)
\end{equation}
where $v(r)$ is the pair interaction potential. It results from Eq. (\ref{eq:pairwise_energy}) that systems with identical
$S_2(r)$ necessarily have the same energy. Therefore the degeneracy $\Omega_0$ calculated from $S_2(r)$ is a lower bound for
the physical ground-state degeneracy of any system with pairwise interaction energy. This includes systems such as frustrated
Ising models for which the ground-state degeneracy is not trivial \cite{Parisi:2006}. Another fascinating field of
application is that of quasi-crystalline microstructures \cite{Levine:1984}, the degeneracy of which could be analyzed
with the general results obtained in the present study.

>From a methodological point of view, the general approach we have developed may be valuable in the manifold of fields where
complex energy landscapes have to be characterized. These include protein folding \cite{Chavez:2004}, complex chemical
reactions \cite{Wales:1998}, phase equilibria in disordered porous materials \cite{Puibasset:2010,Gommes:2012B}, and glass
transitions \cite{Debenedetti:2001}. We hope to investigate some of these aspects in future work.

\begin{acknowledgments}
C.J.G. acknowledges support from the Fonds de la Recherche Scientifique (F.R.S.-FNRS, Belgium) and from the Patrimoine
de l'Universit\'e  de Li\`ege; S.T. and Y.J. were supported by the Office of Basic Energy Science, Division of Materials
 Science and Engineering under Award DE-FG02-04-ER46108.
\end{acknowledgments}

\end{document}